\documentclass[twocolumn,trackchanges]{aastex62}

\usepackage{amsmath}
\usepackage{amssymb}
\usepackage{empheq}
\usepackage{mathrsfs}
				
\usepackage{amssymb}

\newcommand{\G}{\mathcal{G}}
\newcommand{\M}{M_{\star}}

\newcommand{\appropto}{\mathrel{\vcenter{
  \offinterlineskip\halign{\hfil$##$\cr
    \propto\cr\noalign{\kern2pt}\sim\cr\noalign{\kern-2pt}}}}}

\newcommand{\be}{\begin{equation}}
\newcommand{\ee}{\end{equation}}

\def\lta{\,\raise 0.3 ex\hbox{$ < $}\kern -0.75 em \lower 0.7 ex\hbox{$\sim$}\,}
\def\gta{\,\raise 0.3 ex\hbox{$ > $}\kern -0.75 em \lower 0.7 ex\hbox{$\sim$}\,}

\begin{document}


\title{Interactions Among Non-Interacting Particles in Planet Formation Simulations}

\author{Shirui Peng}
\affiliation{Division of Geological and Planetary Sciences, California Institute of Technology, Pasadena, CA 91125, USA}

\author{Konstantin Batygin}
\affiliation{Division of Geological and Planetary Sciences, California Institute of Technology, Pasadena, CA 91125, USA}

\begin{abstract}
Over the course of the recent decades, $N$-body simulations have become a standard tool for quantifying the gravitational perturbations that ensue in planet-forming disks. Within the context of such simulations, massive non-central bodies are routinely classified into ``big" and ``small" particles, where big objects interact with all other objects self-consistently, while small bodies interact with big bodies but not with each other. Importantly, this grouping translates to an approximation scheme where the orbital evolution of small bodies is dictated entirely by the dynamics of the big bodies, yielding considerable computational advantages with little added cost in terms of astrophysical accuracy. Here we point out, however, that this scheme can also yield spurious dynamical behaviour, where even in absence of big bodies within a simulation, indirect coupling among small bodies can lead to excitation of the constituent ``non-interacting" orbits. We demonstrate this self-stirring by carrying out a sequence of numerical experiments, and confirm that this effect is largely independent of the time-step or the employed integration algorithm. Furthermore, adopting the growth of angular momentum deficit as a proxy for dynamical excitation, we explore its dependence on time, the cumulative mass of the system, as well as the total number of particles present in the simulation. Finally, we examine the degree of such indirect excitation within the context of conventional terrestrial planet formation calculations, and conclude that although some level of caution may be warranted, this effect plays a negligible role in driving the simulated dynamical evolution.
\end{abstract}


\section{Introduction}




The past three decades have seen staggering advances in computation, and few sub-fields of astrophysics have benefited from these developments as much as the study of planet formation. Having remained elusive for centuries, the chaotic evolution inherent to the coalescence of planetary building blocks into \textit{bona fide} planets is now within reach of modern GHz-grade machines \citep{1998AJ...116..2067D,1999MNRAS.304..793C,2012A&A...537A.128R}. It is with the detailed numerical modeling of this process that we will concern ourselves in this letter. In particular, here we point out that the conventional approach to modeling quasi-Keplerian, large-$N$-body systems is susceptible to spurious excitation of the constituent orbits, although we find that this effect is negligibly small within real astrophysical applications. Let us begin by briefly outlining the context of our calculations. 

Crudely speaking, the process of planet formation can be sub-divided into two temporal scales: the disk-bearing phase (during which the central star is encircled by an extensive disk of gas and dust) and the post-nebular epoch \citep{2012AREPS..40..251M,Lissauer1993,2013apf..book.....A} which takes place after the large-scale depletion of H and He from the system. In terms of governing physics, the former is subject to a multitude of complex gravito-hydrodynamic processes, while the latter is governed primarily by purely gravitational dynamics \citep[see for example,][for a recent review]{2020arXiv200205756R}. For definiteness, here we will restrict ourselves to consideration of the latter, more qualitatively simple mode of planet formation (corresponding to the post-nebular epoch), where the buildup of planetary bodies proceeds primarily via pair-wise collisions among planetesimals.

Modeling of post-nebular evolution of planetary systems is typically carried out by breaking up the calculation into three types of constituents: big objects, small bodies, and test particles. Big objects interact with all other bodies in a self-consistent $N$-body fashion. Small bodies (sometimes also called semi-active particles) interact with big bodies but not with each other. Finally, test particles merely track the dynamics facilitated by the gravitational field of the big bodies, exerting no back-reaction. The reason for this categorization is two-fold. First, without this treatment, the computational burden of a typical $N$-body simulation would scale as $\mathcal{O}(N^2)$, with $N$ being the total number of bodies in the problem. The big-small categorization, however, alters the computational cost of the simulation to $\mathcal{O}(N_b^2) + \mathcal{O}(N_b N_s)$, with $N_b$ and $N_s$ being the number of big and small bodies, respectively. Because in real planet-formation systems $N_b\ll N_s \sim \mathcal{O}(N)$, this treatment translates to a drastic reduction of computational costs.

Second, ignoring self-gravitational stirring among small bodies circumvents unphysical excitation of the orbits. This is because in an effort to keep $N_s$ to a reasonably low number, the population of solid debris is routinely modeled as a swarm of “super-particles” — objects that trace the dynamics of planetesimals but contain much more mass than the individual bodies they represent. In turn, suppression of self-interactions within the planetesimal swarm prevents the Safronov number $\Theta=(v_{\rm{esc}}/\langle v \rangle)^2$ — which regulates the efficiency of accretion \citep[see e.g.,][]{1972epcf.book.....S,Lissauer1993} — from decreasing artificially. In other words, the big-small particle characterization mimics the un-modeled effect of dynamical friction.


Owing to these advantages, the classification of non-central bodies into big and small particles is widely utilized, with important examples set within the solar system itself. In particular, over the last decade, conglomeration of Mercury, Venus, Earth, and Mars has been modeled by various groups as the gravitational evolution of a $\sim 2 M_{\earth}$ annulus of debris where the initial mass-fraction of “big” planetary embryos to “small“ planetesimals is taken to be approximately unity (\citealt{2014RSPTA.37230174J,2011Natur.475..206W}; see also \citealt{2009ApJ...703.1131H}). Within the context of the outer solar system, a transient (Nice-model) instability is believed to have been sparked early in the solar system’s life time by interactions among “big” planets and a $\sim 20 M_{\earth}$ disk of “small” planetesimals \citep{2005Natur.435..459T,2012AJ....144..117N}. Similarly, recent simulations of the formation of Galilean moons \citep{2020ApJ...894..143B} treat the satellite seeds as “big” objects, while modeling the much more numerous aggregate of satellitesimals as small bodies.

In this paper, we show that even if direct interactions are turned off, some degree of self-stirring within the disk may be unavoidable. More specifically, we carry out tests with different disk to star mass ratios and varying numbers of non-interacting planetesimals. The results generally show a growing trend of angular momentum deficit, indicating a gradual increase of average eccentricity in time. Numerical tests using finer time steps or integrators with higher accuracy give essentially identical results. Nevertheless, our simulations also show that the disk of debris responsible for the generation of terrestrial planets is not sufficiently massive for this effect to meterialize in any appreciable manner.

The remainder of this letter is organized as follows: Section \ref{section:numerics} provides a description of the setup of our numerical experiments. Section \ref{section:rslts} illustrates the key results of our simulations. In Section \ref{section:eg}, the formation of the terrestrial planets is examined as an illustrative example. Our findings are summarized in section \ref{section:sum}. 

\section{Numerical Experiments} \label{section:numerics}

Swarms of planet-forming debris are routinely envisioned to emerge from their natal protoplanetary nebulae, possessing negligible eccentricities and inclinations. Evolving under self-gravitation over timescales much longer than an orbit, massive objects perturb one another, causing the effective velocity dispersion of the system to increase. However, this process is markedly not uniform, as dynamical friction is exerted upon the more massive objects by less massive bodies, causing the largest members of the planet-forming aggregate to circularize at the expense of further excitation of their smaller, but more numerous counterparts \citep{1972epcf.book.....S,Lissauer1993}.

In the language of standard $N$-body simulations, this picture can be summarized in a straight-forward manner: in an initially quiescent disk, big bodies heat each other as well as the small bodies, while small bodies only cool the big bodies. Correspondingly, if no big bodies are present in the simulation, the perfectly circular and coplanar architecture of the system should be preserved. Let us check this assertion with the aid of a state-of-the-art $N$-body code \uppercase{REBOUND} \citep{2012A&A...537A.128R}.

The basic setup of our numerical experiments draws upon standard planet formation simulations. In our fiducial model (M3 in Table~\ref{tab:table}), we represent the disk of planetesimals in orbit of a single central body of mass $\M$ as $N_s=1000$ super-particles, which cumulatively comprise a disk with mass $M_{\mathrm{disk}}/M_{\star}=10^{-4}$, confined to a radius range between $0.1$ and $1$ length units. In our unit system, we set $\M=1$, $\G=1$, and the time and space units are chosen such that an orbit with $1$ semi-major axis unit has a period of $T=2\pi~ \rm{time~ units}$, which we define as a single ``year". The radius of each body is set to zero to suppress any collisions. The semi-major axes are spread within the disk uniformly from $0.1$ to $1.0$, and all orbits are assumed to be initially circular with zero inclination. The default time step is taken to be $1/(21+\varepsilon)$ of the orbital period with a semi-major axis of $0.1$ length units, where $\varepsilon=10^{-3}$ is an arbitrary small quantity. The baseline $N$-body integrator is the hyrbid symplectic intergator \texttt{MERCURIUS} \citep{2019MNRAS.485.5490R}, based upon the widely-used \textit{Mercury6} software package \citep{1999MNRAS.304..793C}. By default, the systems are run for $6 \times 10^5$ time units ($\sim 0.1~\rm{Myr}$).

\begin{deluxetable}{ccccc}
\tablecaption{Model Parameters \label{tab:table}}
\tablehead{
\colhead{Model} & \colhead{$M_{\mathrm{disk}}$} & \colhead{ $N_s$} & \colhead{$\Delta t^{-1}$} & \colhead{ Integrator}\\
\colhead{} & \colhead{$(M_{\star})$} & \colhead{} & \colhead{$(T_{min}^{-1})$}  & \colhead{}
}
\colnumbers
\startdata
M1 & $10^{-3}$ & $1000$ & $21$ & \texttt{MERCURIUS}\\
M2 & $5\times10^{-3}$ & $1000$ & $21$ & \texttt{MERCURIUS}\\
M3 & $10^{-4}$ & $1000$ & $21$ & \texttt{MERCURIUS}\\
M4 & $5\times10^{-4}$ & $1000$ & $21$ & \texttt{MERCURIUS}\\
M5 & $10^{-5}$ & $1000$ & $21$ & \texttt{MERCURIUS}\\
\hline
M6 & $10^{-3}$ & $100$ & $21$ & \texttt{MERCURIUS}\\
M7 & $5\times10^{-3}$ & $100$ & $21$ & \texttt{MERCURIUS}\\
M8 & $10^{-4}$ & $100$ & $21$ & \texttt{MERCURIUS} \\
M9 & $5\times10^{-4}$ & $100$ & $21$ & \texttt{MERCURIUS}\\
M10 & $10^{-5}$ & $100$ & $21$ & \texttt{MERCURIUS}\\
\hline
M11 & $10^{-3}$ & $10000$ & $21$ & \texttt{MERCURIUS}\\
M12 & $5\times10^{-3}$ & $10000$ & $21$ & \texttt{MERCURIUS}\\
M13 & $10^{-4}$ & $10000$ & $21$ & \texttt{MERCURIUS}\\
M14 & $5\times10^{-4}$ & $10000$ & $21$ & \texttt{MERCURIUS}\\
M15 & $10^{-5}$ & $10000$ & $21$ & \texttt{MERCURIUS}\\
\hline
M16 & $10^{-4}$ & $1000$ & $210$ & \texttt{MERCURIUS}\\
M17 & $10^{-4}$ & $1000$ & $21$ & \texttt{IAS-15}\\
M18 & $10^{-4}$ & $1000$ & $21$ & \texttt{EOS}\\
M19 & $10^{-4}$ & $1000$ & $21$ & \texttt{JANUS}\\
M20 & $10^{-4}$ & $1000$ & $21$ & \texttt{LEAPFROG}\\
\enddata
\tablenotetext{}{A summary of simulations carried out in this work. The default model is M3. In the column of $\Delta t^{-1}$, a small correction of $\varepsilon T_{min}^{-1}=10^{-3} T_{min}^{-1}$ is dropped. Our baseline integrator is \texttt{MERCURIUS}, which employs the \citet{1991AJ....102.1528W} mapping. \texttt{IAS-15} is a high accuracy non-symplectic integrator with adaptive timestepping \citep{2015MNRAS.446.1424R}; \texttt{EOS} corresponds to the Embedded Operator Splitting Methods; \texttt{JANUS} is an integer based integrator \citep{2018MNRAS.473.3351R}; and \texttt{LEAPFROG} is a symplectic integrator that does not require a Kepler solver.}
\end{deluxetable}

Besides the fiducial model, we perform a series of tests for different values of $M_{\mathrm{disk}}$ and $N_s$. We also test the robustness of this self-string for different time steps and integrators. A representative list of models is summarized in Table \ref{tab:table}.

In order to avoid examining the dynamics of each simulated particle individually, we quantify the results of our simulations in terms of the angular momentum deficit (AMD) \citep{1997A&A...317L..75L,2000PhRvL..84.3240L}:
\be
\mathrm{AMD}=\sum_{j=1}^N \left(L_j - G_j \right) = \sum_{j=1}^N L_j \left(1-\sqrt{1-e_j^2} \right),
\label{AMD} 
\ee
where $L_j = \mu_j \sqrt{\G (\M+m_j) a_j}$ and $G_j=L_j \sqrt{1-e_j^2}$. In our situation, the reduced mass $\mu_j\approx M_{\mathrm{disk}}/N_s\ll \M$, and $L_j \approx M_{\mathrm{disk}} \sqrt{\G \M a_j}/N_s$. Importantly, low values of this quantity (\ref{AMD}) correspond to near-circular orbits, while high eccentricities give large AMD and small $\Theta$, implying unfavorable conditions for planet formation.

\begin{figure*}[tbp]
\centering
\includegraphics[width=\textwidth]{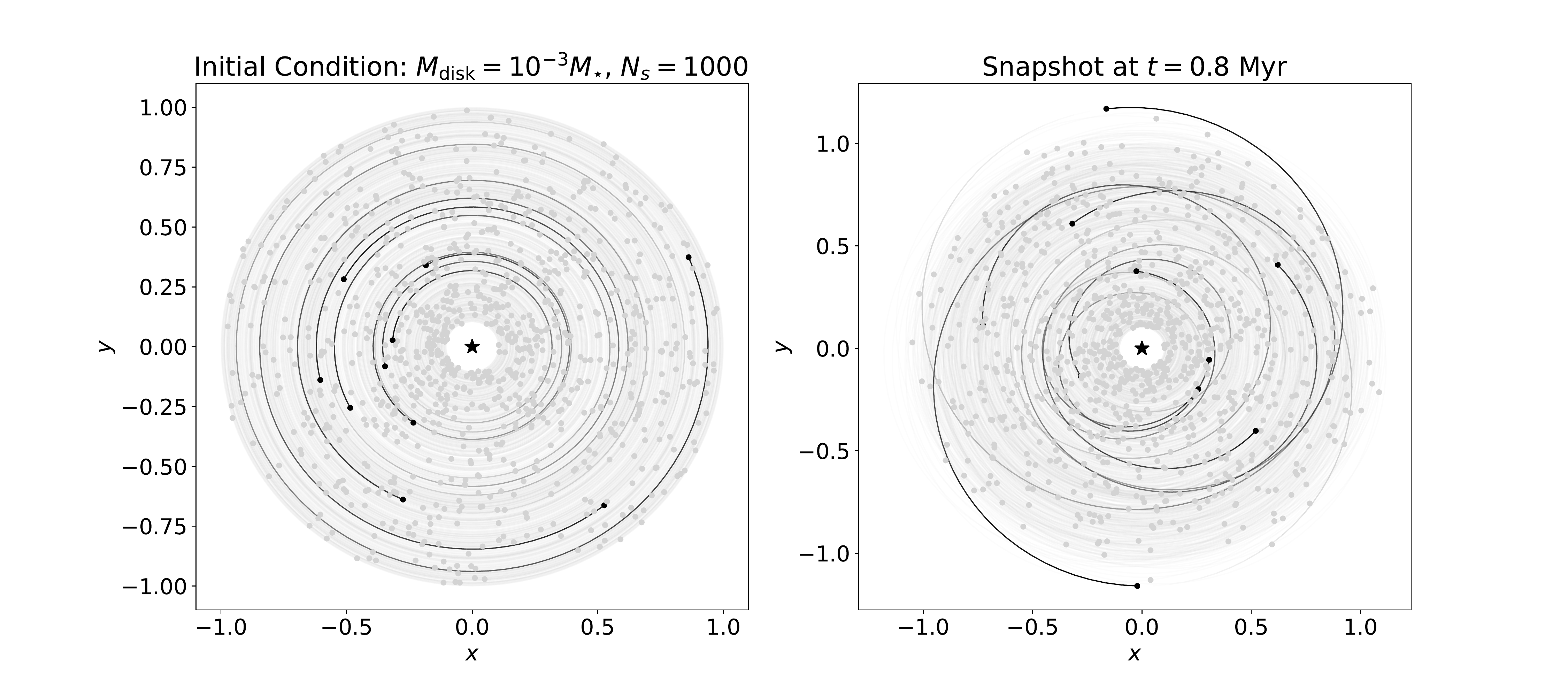}
\caption{The initial (left) and final (right) states of model M1. The small bodies start with perfectly circular orbits, and end with overlapped eccentric orbits, despite the fact that direct gravitational interactions among the particles are suppressed. $990$ of the small bodies and their orbits are displayed in lightgray, and $10$ in black with orbits whose eccentricities evolve above $99^{\rm{th}}$ percentile in the final stage. For the purposes of this example, the integration time was extended to $\sim 0.8~\rm{Myr}$, to better illustrate the effect of self-stirring.}
\label{F:test1}
\end{figure*} 
\section{Results} \label{section:rslts}

\begin{figure*}[tbp]
\centering
\includegraphics[width=\textwidth]{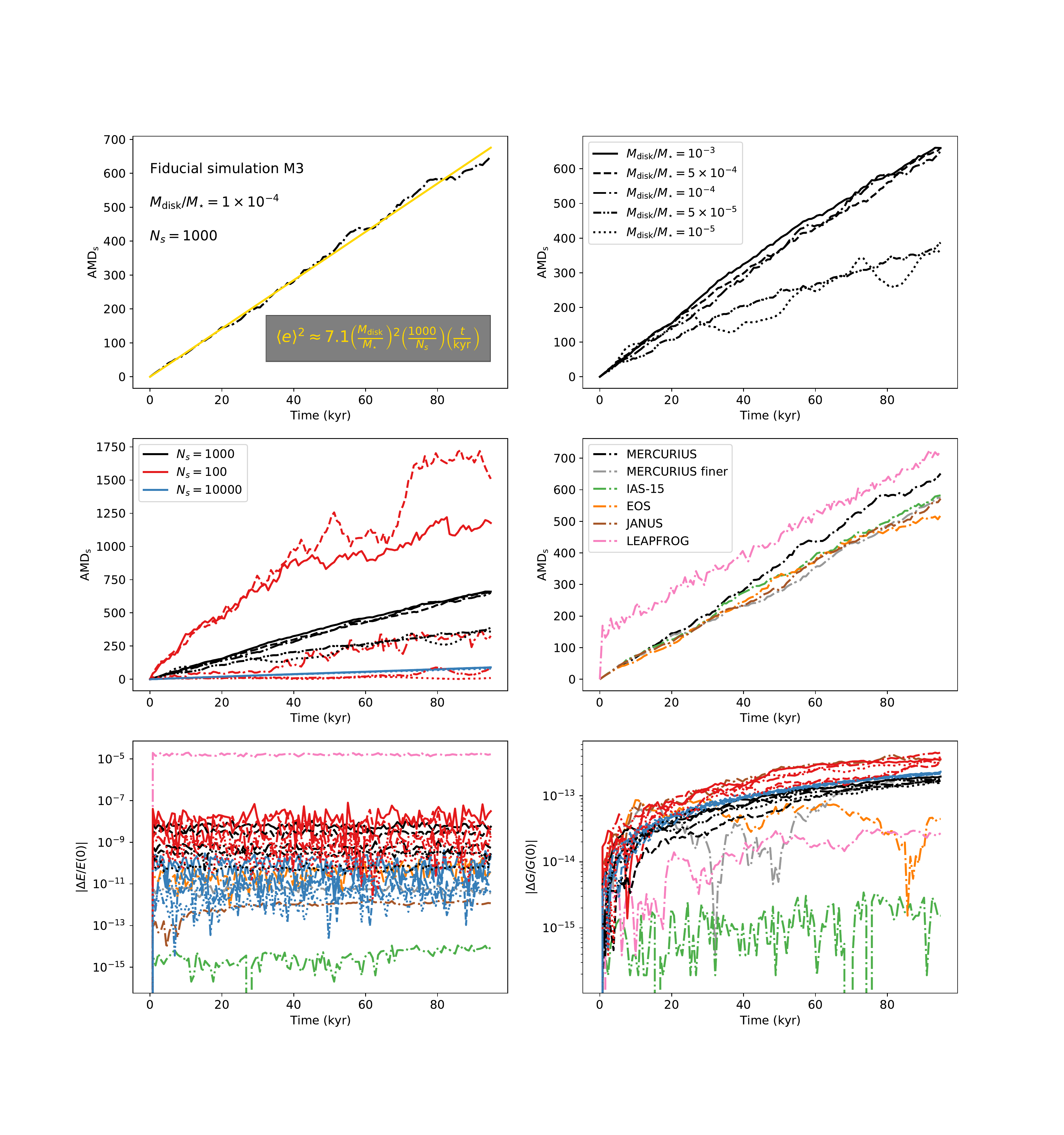}
\caption{Results of our numerical experiments. The time series correspond to the models listed in Table \ref{tab:table}. Different line styles indicate different $M_{\mathrm{disk}}/\M$, and different line colors indicate analytic approximation, different $N_s$ or numerical setups. The first two rows correspond to dependencies on time (upper left), mass (upper right), $N_s$ (middle left), and numerical setup (middle right). The third row shows energy (left) and angular momentum (right) errors.}
\label{F:rslts}
\end{figure*} 

Let us begin with an illustrative example. That is, while all of our models display a certain degree of self-stirring among planetesimals in the disk, the tendency towards self-excitation is more pronounced in more massive disks. Correspondingly, as a demonstration of the described collective behaviour, in Figure (\ref{F:test1}) we show an initial as well as an evolved orbital states of a $M_{\mathrm{disk}}=10^{-3}\M$, $N_s=1000$ disk. More specifically, Figure (\ref{F:test1}) depicts the starting state of the disk with purely circular orbits on the left panel and a final state with overlapped eccentric orbits on the right panel, where the default integration timescale was increased by a factor of eight to accentuate the growth of eccentricities. 

Of course, disks of solid debris as massive as that considered in Figure (\ref{F:test1}) are unlikely to be physical within the broader context of planet formation. Given that gravitational stability limits the mass of quasi-Keplerian disks from above to a value smaller than their aspect ratio $M_{\mathrm{disk}}/M_{\star}\lesssim h/r \sim 0.05$ and that the typical dust-to-gas ratio of circumstellar nebulae is of order $1\%$, we adopt $M_{\mathrm{disk}}/M_{\star}=10^{-4}$ as a reasonable mass-scale for our fiducial experiment, M3. The scaled AMD evolution (see below) obtained within this simulation is shown in the top left panel of Figure (\ref{F:rslts}) as a black curve with dash-dotted line.

\paragraph{Dependence on Time} An immediately notable feature of the depicted time-series is that the growth of the scaled AMD is approximately linear. In fact, such behavior can be expected if the individual eccentricities themselves increase due to stochastic forcing \citep[see e.g.,][]{2018AJ....155..157P}. This can be understood as follows. First, we note that for small eccentricities, the simplification $1-\sqrt{1-e^2}\approx e^2/2$ holds for the rhs of equation (\ref{AMD}). Second, we suppose that that the progress of $e$ is ``diffusive" \citep[see e.g.,][]{2003sde..book.....O} such that $\mathrm{d}e\appropto \mathrm{d}\mathcal{W}$ (Wiener process) yielding $e\appropto\sqrt{t}$. This presumption immediately gives $\mathrm{AMD}\appropto e^2 \appropto t$ for small eccentricities.


\paragraph{Dependence on Mass} The dependence of AMD growth upon disk mass can be reasoned out in a similar fashion. Particularly, if we postulate that $\mathrm{d}e/\mathrm{d}t \propto M_{\mathrm{disk}}/\M$, then $e^2 \appropto (M_{\mathrm{disk}}/\M)^2 \,t$ and  a reasonable choice of AMD. evolution scaling would be $\mathrm{AMD}\appropto M_{\mathrm{disk}}\, e^2\appropto M_{\mathrm{disk}} (M_{\mathrm{disk}}/\M)^2 t$.
If correct, then by defining the scaled AMD as
\be
\mathrm{AMD_s}= \frac{\mathrm{AMD}}{G(0)}\left(\frac{\M}{M_{\mathrm{disk}}}\right)^2,
\label{AMD_s} 
\ee
it should be possible to collapse the time evolution of models with identical $N_s$ but with different $M_{\mathrm{disk}}$ onto a common curve. In other words, to remove the envisioned cubic dependence of AMD growth on $M_{\mathrm{disk}}$, we divide the AMD by the initial angular momentum $G(0)=\sum_{j=0}^N G_j(t=0) \propto M_{\mathrm{disk}}$, and further scale it by the square of disk-to-star mass ratio, rendering $\mathrm{AMD_s}$ dimensionless. 

To test this assertion, we show $\mathrm{AMD_s}$ growth for a series of $N_s=1000$ models (Figure \ref{F:rslts} top right panel), spanning $M_{\mathrm{disk}}=10^{-5}-10^{-3}\M$ (M1 -- M5). The individual numerical experiments are marked with different line styles. By and large, these $N_s=1000$ simulations exhibit approximately linear growth in scaled AMD, and the depicted curves have comparable slopes (to within a factor of $\sim 2$). This means that the mass scaling proposed above is satisfactory, although certainly not exact. More specifically, the three most massive disks (M1, M2, M3) tend to have indistinguishable $\mathrm{AMD_s}$ growth. Interestingly, the time series of the experiments with two smallest disk masses (M4, M5) also tend to overlap fairly well, but have growth rates that are notably smaller. Although of some interest, chasing down the associated correction to equation (\ref{AMD_s}) is beyond the scope of our exploratory paper. Instead, we now turn our attention to the dependence of this effect upon the ``resolution" of our experiments, $N_s$. 

\paragraph{Dependence on $N_s$} Although $N_s\sim1000$ is a routinely adopted particle count in simulations of planet-forming disks, studies employing an order of magnitude more (or fewer) particles are not uncommon in the literature \citep[e.g.,][and the references therein]{2007AmJPh..75..139A,2012Icar..220..777R,2009Icar..203..233C,2016MNRAS.457L..89M,2016ApJ...825...94N}. Correspondingly, we have repeated the aforementioned numerical experiments with $N_s=100$ (M6 -- M10) and $N_s=10,000$ (M11 -- M15), which are shown as red and blue curves in the middle left panel of Figure (\ref{F:rslts}), respectively. The qualitative features of the obtained time-series are readily summarized: simulations with $N_s=100$ exhibit more rapid and more uneven $\mathrm{AMD_s}$ growth than their higher-$N_s$ counterparts. To this end, the concavity of the highest-$\mathrm{AMD_s}$ $N_s=100$ models can likely be attributed to the fact that the attained eccentricities are so high that the reasoning behind quasi-linear growth outlined above no longer applies. Conversely, curves corresponding to $N_s=10,000$ are smooth, linear, and overlap one-another very well, implying that equation (\ref{AMD_s}) constitutes a better approximation for simulations with higher $N_s$. 

In addition to the aforementioned experiments, we have also measured the characteristic Lyapunov timescale of $N_s=1000$ experiments and found that it decreases approximately as the inverse square root of the disk mass. Specifically, for a $M_{\mathrm{disk}}=10^{-5}\M$ system, $T_l \sim 250~\rm{yr}$; for $M_{\mathrm{disk}}=10^{-4}\M$, $T_l \sim 80~\rm{yr}$; and for $M_{\mathrm{disk}}=10^{-3}\M$, $T_l \sim 20~\rm{yr}$. Our $M_{\mathrm{disk}}=10^{-4}\M$ simulations further indicate that the Lyapunov timescale exhibits ancillary dependence on the particle count, with $N_s=100$, $N_s=10000$ runs yielding $T_l \sim 250~\rm{yr}$ and $T_l \sim 130~\rm{yr}$, respectively.

Cumulatively, the results of these experiments are consistent with an interpretation wherein the spurious growth of the angular momentum deficit is driven by perturbations that small particles exert upon the central body, which are then transmitted to other members of the system. In other words, the gravitational coupling we observe in our numerical experiments is likely facilitated in full via the indirect terms of the disturbing Hamiltonian \citep[see Ch. 6 of][]{1999ssd..book.....M}, since there are no other interaction terms in the code. It further worth noting that all indirect terms of the disturbing function average out to zero in the secular limit (where perturbations are taken to be phase-averaged), and indeed, this is the limit we approach as $N_s \rightarrow\infty$, which explains why the rate of $\mathrm{AMD_s}$ growth diminishes with increasing $N_s$. Specifically, we found that $\mathrm{AMD_s}\appropto N_s^{-1}$ via regression for $N_s \sim 1000 - 10000$.

\begin{figure*}[tbp]
\centering
\includegraphics[width=\textwidth]{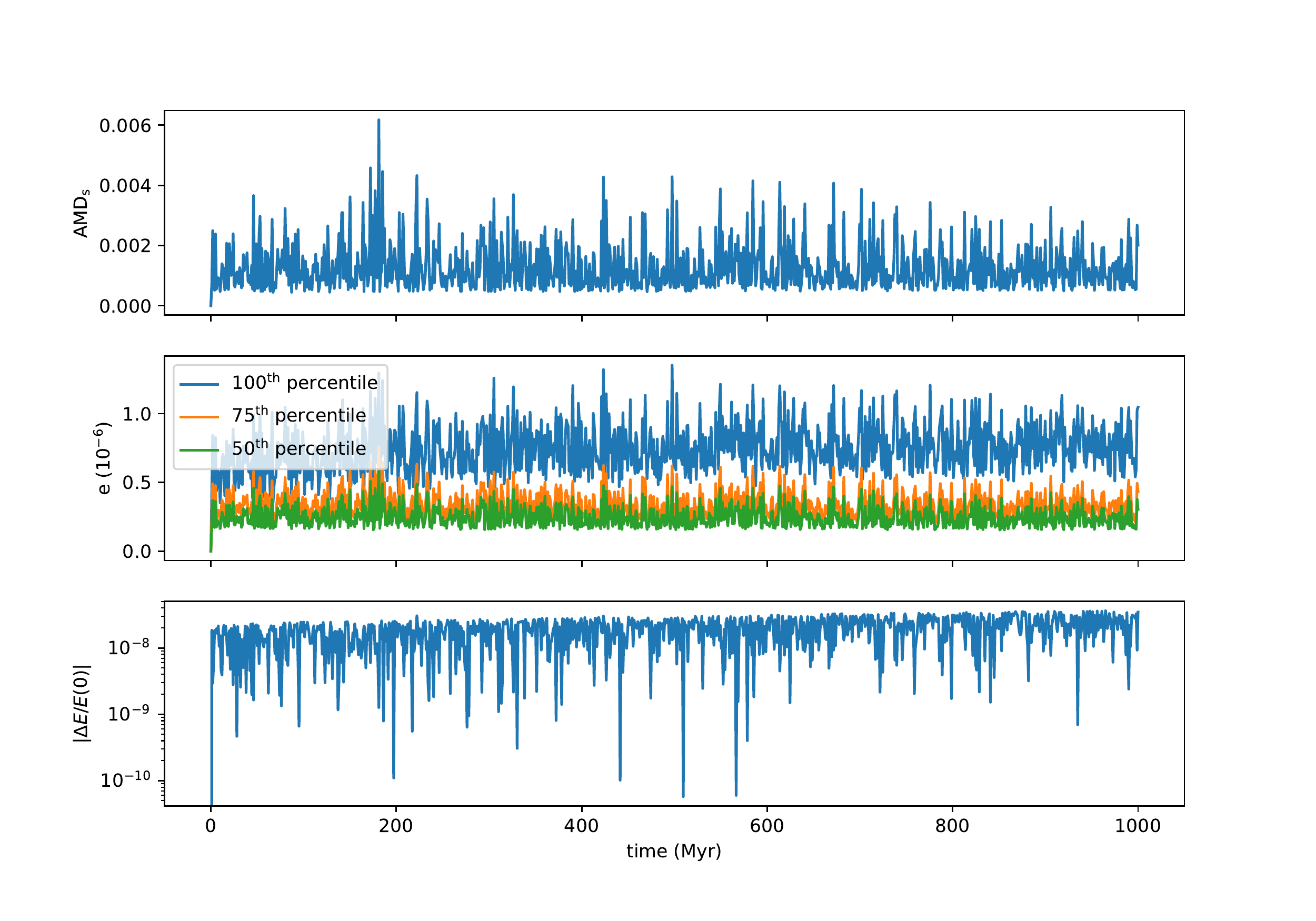}
\caption{Evolution of terrestial planet forming annulus of debris, with direct interactions suppressed. The upper panel shows the evolution of $\mathrm{AMD_s}$ with no clear growth. The middle panel shows the evolution of three percentiles ($100^{\rm{th}}$, $75^{\rm{th}}$ and $50^{\rm{th}}$) in the eccentricity distribution. The lower panel gives the energy error of the integration.} 
\label{F:eg}
\end{figure*} 

\paragraph{Dependence on Timestep \& Integration Method} While the aforementioned dependencies of the collective disk behavior on $t$, $M_{\mathrm{disk}}$, and $N_s$ appear sensible, the above discussion falls short of addressing the possibility that the dynamical behavior observed in our simulations is nothing more than a numerical artifact. Thus, as a final check on our results, we have repeated our fiducial $M_{\mathrm{disk}}=10^{-4}\M$, $N_s=1000$ simulation employing a variety of numerical setups. Specifically, we test the dependence of the observed behavior on timestep (M16, M17), as well as integration method (M17 -- M20). The growth of $\mathrm{AMD_s}$ is depicted on the middle right panel of Figure (\ref{F:rslts}). Importantly, all of these numerical experiments yield consistent results, insinuating that the observed dynamical excitation is genuine, and is not a feature of any specific algorithm. We have further used the \textit{Mercury6} software package \citep{1999MNRAS.304..793C} to reproduce some of our results using the “hybrid” and Bulirsch–Stoer algorithms \citep{1992nrfa.book.....P}, as well as the \texttt{IAS-15} integrator to verify the dependence on $M_{\mathrm{disk}}$ illustrated in the left panel and got good agreement in all cases.

\section{A Heuristic Example} \label{section:eg}

Having demonstrated that indirect gravitational coupling among ``small" non-interacting particles is a generic feature of $N$-body simulations, we are now in a position to inquire if this effect is of appreciable practical importance in real planet-formation calculations. To answer this question, we proceed by considering a specific example of post-nebular dynamical evolution already mentioned in the introduction: the assembly of terrestrial planets from a narrow annulus of rocky debris. In particular, we follow  \citet{2009ApJ...703.1131H} and build our terrestrial planet formation experiment by initializing a $M_{\mathrm{disk}}=6\times10^{-6}\M$ disk of planetesimals confined between $0.7$ and $1$ length units in radial direction, broken up into $1000$ equal mass bodies. The semi-major axes are taken to be spread randomly within the annulus and all orbits are assumed to be initially circular and coplanar. Finally, the time step is taken to be $\sim 5\%$ of the orbital period with a semi-major axis $0.7$ length units, and the integration is run for $2\pi \times 10^9$ time units. 

The relevant time-series of this simulation are summarized in Figure (\ref{F:eg}). Intriguingly, these results show no sustained self-stirring in the system. Instead, contrary to the numerical experiments reported in the previous section, $\mathrm{AMD_s}$ exhibits only low-amplitude fluctuations, and stays below $\mathrm{AMD_s}\lesssim10^{-2}$ (upper panel), as does the eccentricity distribution (middle panel), with $\langle e\rangle  \lesssim 10^{-6}$. We solidify this conclusion by repeating the experiment with different random configurations and alternate integrators, as well as with a smaller number of bodies ($N_s=400$). 

In our interpretation, the disparity between this experiment and those described above lies in that here, $M_{\mathrm{disk}}$ is so low, that the indirect gravitational stirring falls below machine precision. In other words, the effect we describe herein operates only above a threshold mass of the small particle swarm. To test this assertion, we have repeated the \citet{2009ApJ...703.1131H} once again, boosting the disk mass to $M_{\mathrm{disk}}=1\times10^{-3} \M$ as in Figure \ref{F:test1}, and observed growth of the angular momentum that is fully consistent with the results depicted in Figure \ref{F:rslts}. Indeed, it is likely that the precise value of the threshold mass above which indirect self-excitation ensues is both a function of $N_s$ as well as other details of the physical setup of the simulation such as the radial extent of the disk, surface density profile, etc.




\section{Summary} \label{section:sum}

In this work, we have considered the dynamical consequences of big-small particle categorization scheme employed in conventional $N$-body simulations of planet-forming disks. To this end, we have carried out a series of numerical experiments that demonstrate that even in absence of any big non-central bodies, interactions among massive small particles can still yield self-stirring within the system. We argue that this mode of dynamical excitation arises from indirect gravitational coupling, wherein perturbations are transmitted among particles via the barycentric reflex motion of the central star, induced through a superposition of individual Keplerian orbits.

Collectively, our simulation suite shows that the aforementioned effect yields a growth of the system’s angular momentum deficit that is approximately linear in time, and scales roughly as the cube of the cumulative disk mass. These results are consistent with a picture where the evolution of the individual eccentricities is driven by stochastic fluctuations, whose amplitude scales linearly with the disk mass (such that the diffusive progress of the eccentricity dispersion has the approximate form $\langle e \rangle \approx (M_{\mathrm{disk}}/\M) \,(1000/N_s)^{-1/2}\,\sqrt{7.1t/1~\rm{kyr}}$, depicted as the golden line in Figure \ref{F:rslts} upper left panel). Our calculations further show that the obtained results are insensitive to the integration method, but do exhibit significant dependence on the simulation particle count, with large-$N_s$ disks displaying less rapid $\mathrm{AMD_s}$ growth.

Finally, we have examined the role played by spurious excitation of the orbital dispersion within the context of the solar system’s terrestrial planet formation simulations \citep{2009ApJ...703.1131H,2011Natur.475..206W}. Remarkably, we found no sustained growth of the velocity dispersion arising from indirect interactions among small particles, further demonstrating that this effect only operates above a certain threshold mass-scale which the terrestrial planet-forming annulus does not reach. As a result, we conclude that although some caution may be warranted in simulations of massive planetesimal disks, it is unlikely that interactions among non-interacting particles within $N$-body simulations constitute a significant source of uncertainty in numerical models of planetary assembly.



\acknowledgments We are thankful to Matthew J. Holman, Juliette C. Becker, Marguerite Epstein-Martin, Max Goldberg, Tobias Koehne and Elizabeth Bailey for insightful discussions. Additionally, we would like to thank Hanno Rein for providing a thorough and insightful referee report which led to a considerable improvement of the manuscript, as well as his assistance with implementation of numerical experiments. K.B. is grateful to the David and Lucile Packard Foundation and the Alfred P. Sloan Foundation for their generous support. Simulations in this paper made use of the \uppercase{REBOUND} code which is freely available at \url{http://github.com/hannorein/rebound}.

\listofchanges

\end{document}